# TOWARDS A SOLUTION OF THE POLYCYCLIC AROMATIC HYDROCARBON – DIFFUSE INTERSTELLAR BAND HYPOTHESIS

*Xiaofeng Tan (x.tan@jhu.edu)*

**Abstract**. A novel theoretical method is developed to study the polycyclic aromatic hydrocarbon – diffuse interstellar band (PAH-DIB) hypothesis. In this method, a computer program is used to enumerate all PAH molecules with up to a specific number of fused benzene rings. Fast quantum chemical calculations are then employed to calculate the electronic transition energies, oscillator strengths, and rotational constants of these molecules. An electronic database of all PAHs with up to any specific number of benzene rings can be constructed this way. Comparison of the electronic transition energies, oscillator strengths, and rotational band contours of all PAHs in the database with astronomical spectra allows one to identify possible individual PAH carriers of some of the intense narrow DIBs. Using the current database containing up to 10 benzene rings we have selected 8 closed-shell PAHs as possible carriers of the intense $\lambda 6614$ DIB.

**1. Introduction**

The diffuse interstellar bands (DIBs) are absorption features in the near ultraviolet (UV) to near infrared (IR) spectral range seen in the spectra of stars obscured by diffuse interstellar clouds. Since their discovery in 1922,[1] the identification of the carriers of these bands has been a long-standing challenge for scientists of the 20th century. In the past two decades, the hypothesis that free gas-phase PAHs may carry some of the DIBs has become a consensus – which is known as the PAH-DIB hypothesis.[2–7] Pieces of evidence that support this hypothesis include the rich interstellar elemental abundances of carbon and hydrogen, the high photo stability of PAHs, the identification of PAHs in meteorites,[8] and most importantly, the astonishing similarity between the IR signatures of PAHs and the unidentified infrared (UIR) emission bands.[9][10] Unfortunately, due to the insensitivity of IR spectroscopy to the overall structure of the molecule, no individual PAHs can be identified by the comparison of astronomical spectra with IR spectra of PAHs.

To assess the validity of the PAH-DIB hypothesis, electronic spectra of various neutral and ionized PAHs under conditions that mimic those found in the interstellar clouds where the DIBs are originated have been measured. Earlier experimental data were primarily obtained with matrix-isolation spectroscopy (MIS) due to its capability of trapping neutral and ionized PAHs in cryogenic inert-gas solid matrices.[11][12] The spectral features recorded with MIS, however, are shifted and broadened due to the interaction of the trapped species with the solid lattice. This shortcoming was overcome by the application of cavity ring-down spectroscopy (CRDS).[13–15] Only a limited number of PAHs, however, have been studied by CRDS due to the difficulty of bringing these low-vapor pressure molecules into the gas phase in amounts detectable under the laboratory conditions that are relevant for astrophysical studies.[16–24] Notwithstanding all these efforts, no individual PAHs have ever been identified as DIB carriers.

Recent high-resolution astronomical observations have shown partially resolved rotational band contours of several strong DIBs, including the λ5797, λ5850, λ6196, λ6379, and λ6614 DIBs.[25–27] The rotational band contours of these DIBs mount additional strict constraints and provide crucial information on the properties of the carriers of these DIBs. In theory, one can survey the electronic spectra of all PAHs to identify possible matches with the DIB spectra. In practice, however, this approach is not yet possible since only limited number of PAHs have been studied either experimentally or theoretically. As a matter of fact, only several dozens of PAHs with up to 10 fused benzene rings have been studied, which accounts for only ca. 0.1% of all possible PAHs within the size range. Essentially, there are infinite numbers of PAHs out there. But for a certain diffuse band, only limited numbers of PAHs are relevant – the size range of the relevant PAHs can be determined from the rotational band contours. A big question therefore is: "how do we find all relevant PAHs for a certain diffuse band and pre-select good candidates that should be studied first?" In order to answer this question, a systematic method of enumerating all relevant PAHs and calculating their spectroscopic properties (electronic transition energies, oscillator strengths, and rotational constants) is thus highly desirable. In this paper, we present a novel theoretical method – which involves computer enumeration of PAHs, fast calculations of the electronic transition energies, oscillator strengths, and rotational constants of these PAHs, and pre-selection of possible PAH carriers through the search of the generated electronic database of PAHs – to address this difficult problem and to uncover crucial information about the mysterious PAH problem.

This paper is organized as follows: in section 2, a method of enumerating all PAHs with up to a specific number of benzene rings is given. In section 3, the methods for fast calculations of the electronic transition energies, oscillator strengths, and ground-state rotational constants of the generated PAHs are presented. An electronic database of PAHs can be generated this way. In section 4, we pre-select 8 closed-shell PAHs as possible carriers of the λ6614 DIB. Further discussion is given in section 5.

## 2. Computer Enumeration of PAHs

The prerequisite for building a comprehensive database of PAHs is to enumerate all PAHs with a specific number of fused benzene rings or "cells". To do so, one needs an abstract way to represent the carbon skeleton of a PAH. The ideal carbon skeleton of a PAH is referred as a polyhex.[28] By "ideal", we assume all benzene rings in a PAH are identical regular hexagons. Given a polyhex, one can easily add the missing hydrogen atoms to form a PAH molecule and then optimize its geometry by quantum chemical calculations.

Several algorithms exist for enumerating planar polyhexes.[29–33] The method we use is based on the encoding of planar polyhexes in a two-dimensional honeycomb grid. A 4×4 block honeycomb grid is shown in Fig. 1. In this grid, each cell is represented by one grid point located in its center of mass. Each cell can be identified by its unique cell number or equivalently its row and column indices. A polyhex can be encoded by its constituent cells in the honeycomb grid. If two cells are joined together by a common side, we call the two cells or the corresponding two grid points "connected". Each cell can connect to six cells in an indefinite honeycomb grid. A straightforward way of enumerating all planar polyhexes with *h* cells is to start with all polyhexes with *h* – 1 cells and add to each of them a connecting cell to produce polyhexes with *h* cells. This cell-growth approach, however, produces duplicated polyhexes. We use the symmetry property of the honeycomb grid to eliminate enumeration duplications. For an infinite two-dimensional honeycomb grid, there are six rotation operations and six vertical-mirror reflections that transform a cell into another cell. In addition to these twelve rotation and reflection symmetries, there is a translation symmetry which translate a cell in the grid to another. To eliminate duplications, each time a new polyhex with *h* cells is generated, one transforms this polyhex using each of the twelve rotation and reflection operations. The derived twelve polyhexes are then compared to the existing polyhexes with *h* cells. If there exists a polyhex with *h* cells that differs from one of the twelve derived polyhexes by a translation operation, then the newly generated polyhex has already been generated before and is discarded; otherwise, it is a new polyhex. To ensure fast comparison, all enumerated polyhexes are stored in a binary search tree.

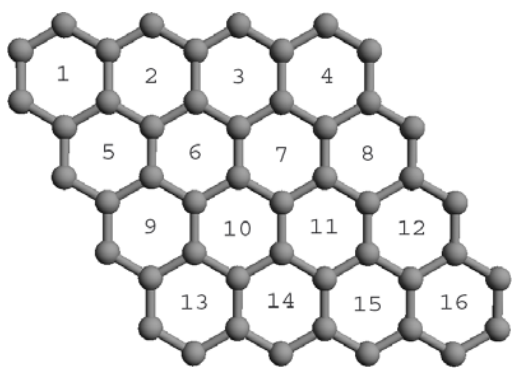

Figure 1. A 4×4 block honeycomb grid. The grid points are not shown on the figure but labeled numerically from left to right and from top to bottom.

As an example, all 7 unique polyhexes with 4 cells are shown in Fig. 2 in which the polyhexes are denoted using their constitution cells in the honeycomb grid – for example, C(1, 2, 3, 4) is the polyhex consisting of cells 1, 2, 3, and 4 in the 4×4 block honeycomb grid as shown in Fig. 1. The fifth polyhex C(1, 2, 3, 6) contains an odd number of carbon atoms while the others contain even numbers of carbon atoms. The neutral PAH of the fifth polyhex is an open-shell radical while its first cation is a closed-shell molecule. Species like this are believed to be significant intermediates in combustion processes and possible interstellar species.[34][35] We list in Table 1 the number of polyhexes with up to 12 cells and the corresponding computational time on an Intel Pentium 4 3.33 GHz computer. It is shown from the Table that the number of polyhexes and the corresponding computational time increase exponentially with the number of cells. This algorithm based on the encoding of honeycomb grid is found to have about the same efficiency as the algorithm based on the Dualist Angle-restricted Spanning Tree code (Nikolić et al 1990).[30]

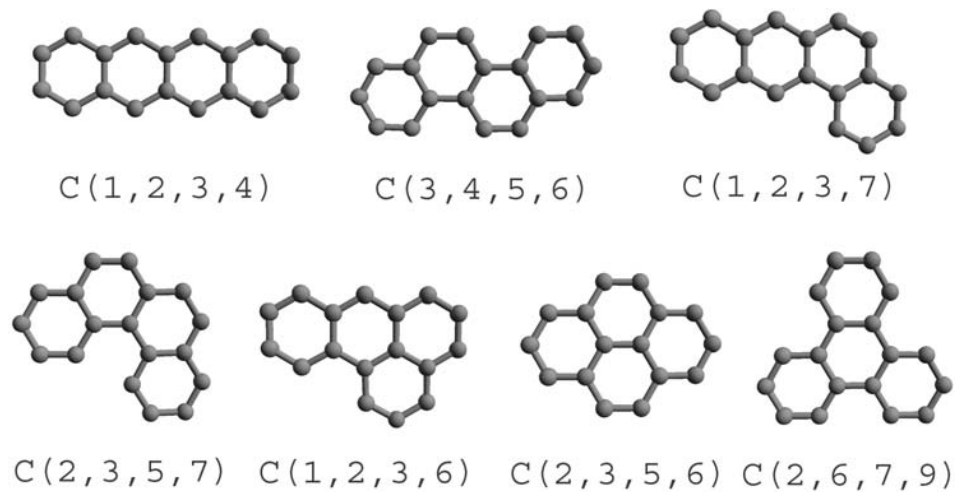

Figure 2. All 7 unique polyhexes with 4 cells.

Table 1. Number of polyhexes and the corresponding computational time.

| h | No. Polyhexes | Time (s) | h | No. Polyhexes | Time (s) |
| --- | --- | --- | --- | --- | --- |
| 1 | 1 | < 1 | 7 | 333 | 1 |
| 2 | 1 | < 1 | 8 | 1448 | 6 |
| 3 | 3 | < 1 | 9 | 6572 | 29 |
| 4 | 7 | < 1 | 10 | 30490 | 167 |
| 5 | 22 | < 1 | 11 | 143552 | 746 |
| 6 | 82 | < 1 | 12 | 683101 | 6527 |

**3. Fast Generation of Electronic Database of PAHs**

The most important step of building a PAH database is to calculate the electronic transition energies, oscillator strengths, and ground-state rotational constants (or equilibrium geometries) of the generated PAHs quickly enough without losing much accuracy. The initial geometries of PAHs can be easily generated from the geometric data of polyhexes. Missing hydrogen atoms are added back to those carbon atoms that are connected with two and only two other carbon atoms. We set the H–C–C angles to $2\pi/3$, the C–C bond lengths to 1.397 Å, and C–H bond lengths to 1.084 Å. These PAH bond lengths are used in the semi-empirical MO/8E force field program.[36] The initial PAH geometries obtained this way are called the "model" geometries. The model geometries are then optimized using quantum chemical methods to obtain the optimized geometries.

Considering the large number of polyhexes (i.e. 38,959 for $h$ up to 10), the methods used for geometry optimization and the calculation of electronic transition energies need to be very efficient. *ab initio* calculations are too computationally expensive for these PAHs and thus not feasible. Instead, semi-empirical methods are more desirable for this purpose. For geometry optimization, we use the Austin Model 1 (AM1) Hamiltonian[37–40] – which is based on the Neglect of Diatomic Differential Overlap (NDDO) approximate Hamiltonian; for electronic excitation energies and oscillator strengths, we use the Zerner's Intermediate Neglect of Diatomic Differential Overlap (ZINDO) method.[41–44] It is found, based on the testing of several PAHs, that the rotational constants derived from the AM1 method are usually 2–5% larger than those obtained from the model geometries. The vertical transition energies calculated with the ZINDO method at the model geometries are usually better than those calculated at the geometries optimized at the AM1 level of theory – presumably due to systematic cancellation of errors for the ZINDO calculations at the model geometries. With these results, we therefore use in the database the ZINDO energies calculated at the model geometries. The rotational constants in the database are still derived from the geometries optimized at the AM1 level of theory.

For calibration purposes, we list in Table 2 the calculated ground-state rotational constants and electronic transition energies of 10 closed-shell neutral PAHs ranging from 1 to 7 benzene rings. We also list available experimental data from gas-phase experiments (18 rotational constants and 15 transition energies) in the Table. It can be seen that the AM1 method predicts very reliable ground state rotational constants that distribute almost evenly around the experimental values. The maximum relative error of the calculated rotational constants is 0.76% and the standard deviation of the relative errors is 0.39%. The electronic transitions energies, however, are all systematically underestimated by the ZINDO method at the model geometries: the mean is $-1059$ cm$^{-1}$ ($-0.13$ eV); the largest deviation is $-2757$ cm$^{-1}$ ($-0.34$ eV); and the standard deviation is 1322 cm$^{-1}$ (0.16 eV).

Table 2. Comparison of calculated and experimentally derived ground-state rotational constants and electronic transition energies of several PAHs in the database (DB). All values are in units of cm$^{-1}$.

| Molecule | No. Rings | Rotational Constants | | Transition Energies | |
|---|---|---|---|---|---|
| | | DB [a] | Expt. | DB [b] | Expt. |
| Benzene ($C_6H_6$) | 1 | $A'' = 0.18972$ | 0.18977 [c] | 52359 | |
| | | $B'' = 0.18972$ | 0.18977 [c] | 47994 | |
| | | $C'' = 0.094861$ | 0.094885 [c] | 37964 | 38606 [c] |
| Naphthalene ($C_{10}H_8$) | 2 | $A'' = 0.10413$ | 0.104385 [d] | 43968 | |
| | | $B'' = 0.041197$ | 0.041513 [d] | 34728 | 35900 [e] |
| | | $C'' = 0.029519$ | 0.029701 [d] | 31425 | 32020 [e] |

| Molecule | | Rotational constants | | Transition energies | |
|---|---|---|---|---|---|
| Anthracene ($C_{14}H_{10}$) | 3 | $A'' = 0.071766$ | $0.07175$ [f] | 35201 | |
| | | $B'' = 0.015127$ | $0.01514$ [f] | 27679 | |
| | | $C'' = 0.012493$ | $0.01250$ [f] | 26166 | 27695 [f] |
| Phenanthrene ($C_{14}H_{10}$) | 3 | $A'' = 0.053814$ | | 38782 | 40000 [g] |
| | | $B'' = 0.018538$ | | 33022 | 35375 [g] |
| | | $C'' = 0.013788$ | | 28894 | 29328 [g] |
| Tetracene ($C_{18}H_{12}$) | 4 | $A'' = 0.054735$ | $0.05437$ [h] | 29595 | |
| | | $B'' = 0.007158$ | $0.007118$ [h] | 25502 | |
| | | $C'' = 0.006330$ | $0.006298$ [h] | 21504 | 22397 [h] |
| Pyrene ($C_{16}H_{10}$) | 4 | $A'' = 0.033822$ | | 32472 | |
| | | $B'' = 0.018569$ | | 28213 | 30970 [i] |
| | | $C'' = 0.011987$ | | 26885 | 27208 [i] |
| Pentacene ($C_{22}H_{14}$) | 5 | $A'' = 0.044246$ | $0.04405$ [j] | 25468 | |
| | | $B'' = 0.003937$ | $0.003935$ [j] | 24155 | |
| | | $C'' = 0.003615$ | $0.003612$ [j] | 18365 | 18649 [j] |
| Perylene ($C_{20}H_{12}$) | 5 | $A'' = 0.020786$ | $0.02068$ [k] | 28610 | |
| | | $B'' = 0.011164$ | $0.01117$ [k] | 27872 | |
| | | $C'' = 0.007263$ | $0.007272$ [k] | 22885 | 24060 [l] |
| Benzo[ghi]perylene ($C_{22}H_{12}$) | 6 | $A'' = 0.014936$ | | 30588 | |
| | | $B'' = 0.011154$ | | 26294 | 27132 [m] |
| | | $C'' = 0.006385$ | | 24874 | 25795 [m] |
| Coronene ($C_{24}H_{12}$) | 7 | $A'' = 0.011160$ | | 32646 | |
| | | $B'' = 0.011160$ | | 27718 | |
| | | $C'' = 0.005580$ | | 23482 | 23822 [n] |

[a] Rotational constants calculated at the AM1 level of theory.
[b] Vertical transition energies of the first three excited singlet states calculated at the ZINDO level of theory at the model geometries.
[c] From Ref. [45]. Transition energy is the origin of the $6_0^1$ band.

<sup>d</sup> From Ref. [46].
<sup>e</sup> From Ref. [47].
<sup>f</sup> From Ref. [48].
<sup>g</sup> From Ref. [49].
<sup>h</sup> From Ref. [50].
<sup>i</sup> From Ref. [51].
<sup>j</sup> From Ref. [52].
<sup>k</sup> From Ref. [53].
<sup>l</sup> From Ref. [22].
<sup>m</sup> From Ref. [23].
<sup>n</sup> From Ref. [54].

The calculated electronic transition energies and ground-state rotational constants will be used to pre-select possible individual PAH carriers of certain intense narrow DIBs. Although the calculated electronic energies are not as accurate as those obtained from high-level *ab initio* methods, they are much more computationally inexpensive to obtain. One can always use a wide energy range in the pre-selection step so that the probability of missing valid candidates is below a specific threshold value.

**4. Pre-selection of Possible Individual PAH Carriers**

With the development of the aforementioned method, we have built an electronic database of PAHs with $h$ up to 10 using a desktop computer. It took us about three weeks to build such a database. More computational resource is needed in order to build a larger database. For example, we estimate that 18 weeks are needed to build a database with $h$ up to 11. However, even with the current database, we can derive some very useful results that cannot be obtained from any other available method. It should be pointed out that a planar polyhex with $h \geq 6$ could represent a PAH with more than $h$ benzene rings. For example, a polyhex with one [6]annulene-like hole represents a PAH with $h + 1$ benzene rings.[55] For this reason, the current PAH database contains not only all PAHs with up to 10 benzene rings but also some PAHs with more benzene rings. The real power of this database approach is that it allows us to survey all PAHs in the database – no PAH is missed within the specific size range of the database – and identify possible matches of the electronic transition energies, oscillator strengths, and rotational contours of PAHs with those of the DIBs. An application of this approach to the identification of possible individual PAH carriers of the λ6614 DIB is demonstrated as follows.

The λ6614 DIB is a strong narrow DIB with an integrated intensity that is ca. 1/8 of that of the strongest λ4427 DIB.[7] High-resolution observations show that this diffuse band has a typical bandwidth of ca. 2 cm$^{-1}$ and three fine-structure peaks with the middle one carrying the strongest intensity. The band contour of this DIB is characteristic of the three main rotational branches of a large molecule. It has been shown that open-shell PAHs have very short-lived excited states with typical lifetimes in the order of 10$^{-13}$ second and produce broader bandwidths (> 15 cm$^{-1}$).[24,17,19,56] If the carrier of the λ6614 DIB is a PAH molecule, then only closed-shell PAHs are possible. Most stable PAHs are planar molecules and in general asymmetric tops except those belong to the D$_{3h}$ or D$_{6h}$ point groups. For these planar PAHs, only two rotational constants are independent since the relation $1/A'' + 1/B'' = 1/C''$ holds. We notice that the rotational constants in the low-lying electronic states of PAHs usually differ from those in the ground state by only a few percent. Therefore, in order to pre-select possible PAH carriers of the λ6614 DIB, we have simulated the rotational band contour of this DIB by assuming that the upper and lower electronic states have the same rotational constants, namely, $A' = A'' = A$, $B' = B'' = B$. We also notice that $\pi \to \pi^*$ or $\sigma \to \sigma^*$ electronic transitions dominate the low-lying electronic transitions of planar PAHs, which produce type-A or type-B rotational band contours. Type-B transitions of these PAHs produce characteristic two-peak rotational contours, which do not match the rotational band contour of the λ6614 DIB. It is thus reasonable to conclude that only type-A or mixed type transitions with a dominant type-A contribution of a closed-shell PAH could possibly produce a rotational band contour similar to that of the λ6614 DIB.

We present in Fig. 3 the λ6614 DIB spectrum obtained from the line of sight towards the object HD149757 by Galazutdinov et al[25] and three band contour simulations of type-A [Fig. 3(a)] and type-B [Fig. 3(b)] transitions. In the simulations, the rotational temperatures are assumed to be 54 K, which is the rotational temperature of $H_2$ along the line of sight towards HD149757;[57] the rotational lines are assumed to have a homogeneous Lorentzian width of 0.4 cm$^{-1}$, which produces the best match with the widths of the three main rotational branches of the λ6614 DIB. In the three simulations, $A' = A'' = A$, $B' = B'' = B$, $A = 0.01$ cm$^{-1}$ and $B/A$ ratios are 0.4, 0.5, and 0.6, respectively. It can be seen from the Fig. 3(b) that type-B transitions of planar PAHs cannot produce the observed band contour of the λ6614 DIB. The simulated rotational band contours are symmetric since the rotational constants in the excited state are assumed to be the same as those in the ground state. These three simulations produce the $B/A$ ratio range within which the simulated rotational band contour reasonably reproduces that of the λ6614 DIB. A smaller $B/A$ ratio than 0.4 produces three bands that are too similar in intensity and too close to each other; a larger $B/A$ ratio than 0.6 produces two side bands that are too broad but too weak in intensity. It should be pointed out that for the same $B/A$ ratio the separations between the three rotational branches are approximately proportional to $(T_{rot} \cdot A)^{1/2}$. One needs to maintain a constant product of the rotational temperature and the rotational constant $A$ in order to maintain a similar band contour for a specific $B/A$ ratio. We have also performed rotational band contour simulations of mixed type-A and type-B transitions and found that within the above $B/A$ ratio range a band ratio above ca. 0.2 will produce contours with significant low-intensity middle peaks.

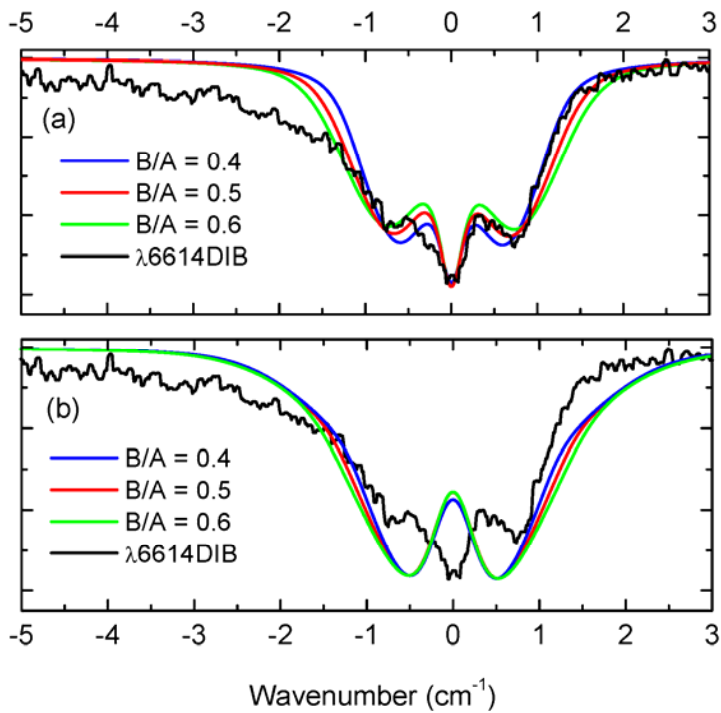

Figure 3. Comparison of the λ6614 DIB (HD149757, Ref. 25) and simulated type-A (a) and type-B (b) transitions of planar PAHs. In the three simulations, $A' = A'' = A$, $B' = B'' = B$, $T_{rot} = 54$ K, $A'' = 0.01$ cm$^{-1}$, $\gamma = 0.4$ cm$^{-1}$. $B/A$ ratios are shown on the figure.

We have surveyed the current PAH database with $h$ up to 10 with the criteria: A = 0.0075–0.019 cm$^{-1}$, which corresponds to $T_{rot}$ = 2.8–72.0 K; B/A = 0.4–0.6; the existence of an electronic transition (referred as the target transition thereafter in the paper) in the range 11416–16704 cm$^{-1}$, which corresponds to the "2σ" 95% confidence interval of 6614 Å in the database – (1E7/661.4 − 1059) ± 2×1322 cm$^{-1}$; and non-existence of an electronic transition with an oscillator strength that is 10 times stronger than the target transition in the range between the target state wavenumber and the lower limit of the λ4427 DIB in the database – (1E7/442.7 − 1059) − 2×1322 cm$^{-1}$ = 18,886 cm$^{-1}$. The last criterion was chosen since we know that in the chosen range there should be no electronic transition that is 10 times stronger than the target state. The database contains closed-shell PAHs – neutral PAHs with even numbers of carbon atoms and the first positively charged PAHs with odd numbers of carbon atoms. It has been shown that PAHs within this size range primarily exist in the ISM as two charge states – neutral molecules and the first cations – due to UV irradiation from nearby stars.[58] Our search returns 64 PAH molecules from the database. Time dependent density functional (TDDFT)[59][60] calculations at the B3LYP/6-31G(d,p)[61–63] level of theory have been performed at the optimized geometries obtained at the AM1 level of theory to further narrow down the candidate list. We discard those PAHs whose electronic transition energies obtained at the B3LYP/6-31G(d,p) level of theory are not within the 570.16 nm – 787.41 nm range or whose type-B to type-A band ratios exceed 0.2. This electronic transition energy range is chosen since it is estimated that the accuracy of the B3LYP/6-31G(d,p) method is within 0.3 eV for calculating electronic transition energies of PAHs.[64] We also discard those PAHs with strong intramolecular forces due to the repulsion between different groups of the molecules – the optimized geometries of these PAHs are usually far away from planar and are unlikely formed in the ISM. Eight PAH molecules – 2 neutral PAH and 6 closed-shell PAH cations ranging from 26 to 33 carbon atoms – are pre-selected this way. Of these 8 PAHs, 2 PAHs contain 7 rings; 4 PAHs contain 8 rings; 2 PAHs contain 9 rings. The 8 pre-selected possible PAH carriers are shown in Fig. 4. The corresponding electronic transitions of these 8 PAHs are listed in Table 3. It has been shown that the carrier of the λ6614 DIB probably has an ionization potential (IP) of 10 to 13 eV.[65][66] This seems to strengthen the candidacy of the 6 pre-selected PAH cations since the IPs of the PAH first cations are usually larger than 10 eV.

We expect no PAHs outside of our current PAH database to be relevant to this λ6614 DIB problem for $T_{rot}$ = 2.8–72.0 K since their rotational constants would be too small to fit the rotational band contour of this DIB within the rotational temperature range – for this particular DIB and the specific rotational temperature range ($T_{rot}$ = 2.8–72.0 K), our current database is complete since it contains all relevant PAHs. To find possible PAH carriers of the λ6614 DIB for a wider rotational temperature range that may be applicable to interstellar PAHs,[67][68] we need to take into account of PAHs with $h \geq 11$ that are not present in the current database.

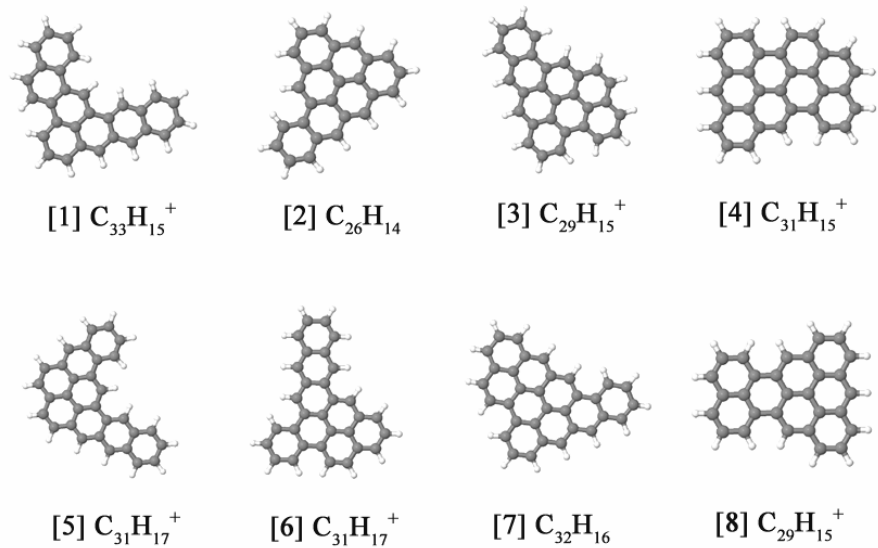

Fig. 4 Eight PAH species – 6 cations and 2 neutral molecules – are pre-selected as possible carriers of the λ6614 DIB.

Further screening of the candidate PAHs is possible. For example, excited-state geometry optimization of these candidates can be performed with the TDDFT method, the Configuration Interaction with Single excitation and Double correction [CIS(D)] method,[69] or the CIS(D) method with the Resolution of Identity approximation [RI-CIS(D)] which allows for efficient computation of the electronic Coulomb interaction and leads to significant speedup for large molecules. This would lead to more accurate rotational band contour simulations and therefore allow us to eliminate more candidates. The excited-state geometry optimizations of these PAHs, however, demand significant amount of computational resource that we do not currently have.

Table 3. The 8 pre-selected PAHs and their corresponding electronic transitions that might be responsible for the λ6614 DIB.

| Molecule | No. Rings | Transition | Wavelength (nm) [a] | Oscillator Strength | Band Ratio [b] |
|---|---|---|---|---|---|
| [1] $C_{33}H_{15}^+$ | 7 | $^1S_2 \leftarrow {}^1S_0$ | 573.3 | 0.25 | 0.12 |
| [2] $C_{26}H_{14}$ | 7 | $^1S_2 \leftarrow {}^1S_0$ | 702.6 | 0.017 | 0.0089 |
| [3] $C_{29}H_{15}^+$ | 8 | $^1S_2 \leftarrow {}^1S_0$ | 617.9 | 0.11 | 0.18 |
| [4] $C_{31}H_{15}^+$ | 9 | $^1S_2 \leftarrow {}^1S_0$ | 703.9 | 0.062 | 0.0038 |
| [5] $C_{31}H_{17}^+$ | 8 | $^1S_1 \leftarrow {}^1S_0$ | 734.2 | 0.23 | 0.030 |
| [6] $C_{31}H_{17}^+$ | 8 | $^1S_2 \leftarrow {}^1S_0$ | 746.2 | 0.17 | 0.19 |
| [7] $C_{32}H_{16}$ | 9 | $^1S_3 \leftarrow {}^1S_0$ | 642.9 | 0.025 | 0.095 |
| [8] $C_{29}H_{15}^+$ | 8 | $^1S_2 \leftarrow {}^1S_0$ | 611.4 | 0.0028 | 0.00 |

[a] Vertical transition energy calculated at the B3LYP/6-31G(d,p) level of theory at the AM1 optimized geometry.
[b] Calculated using the transition dipole moment obtained at the B3LYP/6-31G(d,p) level of theory at the AM1 optimized geometry.

**5. Discussion**

In this paper we present a systematic approach to building an electronic database containing all PAHs with up to any specific $h$ number and using the generated database to pre-select possible individual PAH carriers of certain DIBs. As a test case, 8 closed-shell PAHs have been pre-selected using the current PAH database ($h \leq 10$) as possible carriers of the λ6614 DIB. In general, this approach can also be applied to the identification of possible PAH carriers of other intense narrow DIBs as long as these DIBs are associated with an unique vibronic transition and are caused by closed-shell carriers so that spin-orbit interaction does not play a role in the rotational band contour simulations – the spin-orbit interactions in open-shell PAHs are completely beyond our computational capability. A larger electronic database with $h \geq 11$ is certainly desirable if the DIBs that are narrower than λ6614 DIB (i.e. the λ5797 DIB) and wider rotational temperature range are to be considered. With the availability of more computational resource, a number of narrow DIBs can be studied using this approach and short lists of possible individual PAH carriers of these DIBs can be produced. These candidates can then be studied by further theoretical or experimental techniques to provide crucial information on the validity of the PAH-DIB hypothesis.

Several factors affect the effectiveness of the presented approach. The major one is the practical difficulty of calculating the electronic transition energies and excited-state rotational constants of medium-sized PAHs (with 20 – 50 carbon atoms) accurately enough within reasonable time scale. High level *ab initio* methods, such as the Multi-configuration Self-consistent Field (MCSCF or CAS) and Multi-reference Configuration Interaction (MRCI) are known to be powerful in treating electronic transitions and excited states. However, these methods are far too computationally expensive to be used for medium-sized PAHs till at least the near future. Before then, semi-empirical, TDDFT, and CIS(D) are still the only practicable methods for the presented approach. Another factor that could significantly affect the effectiveness of the present approach is the existence of spectroscopic perturbations in PAHs. The most common form of spectroscopic perturbation in closed-shell PAHs is the vibronic interaction that couples the nuclear motion with the electronic motion of the molecule. The vibronic interaction in PAHs is often not negligible since the existence of great number of nearly degenerate electronic π orbitals. The vibronic interaction affects both the transition energy and the rotational band contour, with a more noticeable effect on the latter since the perturbation on the transition energy is usually much smaller than the error of the calculated electronic transition energy. For example, it has been shown that the vibronic interaction in neutral benzo[ghi]perylene,[23] pyrene,[20,51,70] naphthalene,[71] azulene,[72] phenanthrene,[73] and quinoxaline[74] significantly changes the rotational band contours of these molecules. A common trend of the effect is the broadening of the rotational band contour due to the opening of new decay channels via vibronic interaction and/or the mixing of several closely spaced vibronic levels in the band contour. Without the knowledge of the carrier of a specific DIB, it is usually very difficult to tell if the DIB is subject to vibronic perturbation by simple looking at its band contour. However, by restricting to narrow BIDs, one can more safely ignore the vibronic interaction in the application of the presented approach.

**Acknowledgment**. The author is very grateful to Dr. Jan Cami and Dr. Louis J. Allamandola for the discussion of the properties of the DIBs and the enumeration of PAHs.

# REFERENCES

[1] Heger, M. C. *Lick Observatory Bulletin* **1922**, 10, 146.

[2] Tielens, A. G. G. M.; Snow (Eds.), T. P. *The Diffuse Interstellar Bands*; Kluwer Academic Publ.: Dordrecht, Netherlands, 1995.

[3] van der Zwet, G. P.; Allamandola, L. J. *Astron. Astroph.* **1985**, 146, 76.

[4] Léger, A.; d'Hendecourt, L. B. *Astron. Astroph.* **1985**, 146, 81.

[5] Salama, F.; Bakes, E. L. O.; Allamandola, L. J.; Tielens, A. G. G. M. *Astrophys. J.* **1996**, 458, 621.

[6] Herbig, G. H. *Annu. Rev. Astron. Astrophys.* **1995**, 33, 19.

[7] Tuairisg, S. O.; Cami, J.; Foing, B. H.; Sonnentrucker, P.; Ehrenfreund, P. *Astron. Astrophys. Suppl. Ser.* **2000**, 142, 225.

[8] Ehrenfreund, P.; Charnley, S. B. *Annu. Rev. Astron. Astrophys*. **2000**, 38, 427.

[9] Allamandola, L. J.; Hudgins, D. M.; Sandford, S. A. *Ap. J*. **1999**, 511, L115.

[10] Langhoff, S. R. *J. Phys. Chem*. **1996**, 100, 2819.

[11] Salama, F.; Allamandola, L. J. *J. Chem. Phys*. **1991**, 94, 6964.

[12] Salama, F.; Joblin, C.; Allamandola, L. J. *J. Chem. Phys*. **1994**, 101, 10252.

[13] Berden, G.; Peeters, R.; Meijer, G. *Int. Rev. Phys. Chem*. **2000**, 19, 565.

[14] Herbelin, J. M.; Mckay, J. A.; Kwok, M. A.; Ueunten, R. H.; Urevig, D. S.; Spencer, D. J.; Herbst, D. J. E. *Annu. Rev. Phys. Chem*. **1995**, 46, 27.

[15] Anderson, W. P.; Edwards, W. D.; Zerner, M. C. *Inorganic Chem*. **1986**, 25, 2728.

[16] Romanini, D.; Biennier, L.; Salama, F.; Kachanov, A.; Allamandola, L. J.; Stoeckel, F. *Chem. Phys. Lett*. **1999**, 303, 165.

[17] Biennier, L.; Salama, F.; Allamandola, L. J.; Scherer, J. J. *J. Chem. Phys*. **2003**, 118, 7863.

[18] Biennier, L.; Salama, F.; Gupta, M.; O'Keefe, A. *Chem. Phys. Lett*. **2004**, 387, 287.

[19] Sukhorukov, O.; Staicu, A.; Diegel, E.; Rouillé, G.; Henning, T.; Huisken, F. *Chem. Phys. Lett*. **2004**, 386, 259.

[20] Rouillé, G.; Krasnokutski, S.; Huisken, F.; Henning, T.; Sukhorukov, O.; Staicu, A. *J. Chem. Phys*. **2004**, 120, 6028.

[21] Staicu, A.; Sukhorukov, O.; Rouillé, G.; Henning, T.; Huisken, F. *Mol. Phys*. **2004**, 102, 1777.

[22] Tan, X.; Salama, F. *J. Chem. Phys*. **2005**, 122, 084314.

[23] Tan, X.; Salama, F. *J. Chem. Phys*.**2005**, 123, 014312.

[24] Tan, X.; Salama, F. *Chem. Phys. Lett*. **2006**, 422, 518.

[25] Galazutdinov, G. A.; Moutou, C.; Musaev, F. A.; Krełowski, J. *Astron. Astrophys*. **2002**, 384, 215.

[26] Galazutdinov, G. A.; Musaev, F. A.; Bondar, A. V.; Krełowski, J. *Mon. Not. R. Astron. Soc*. **2003**, 345, 365.

[27] Kerr, T. H.; Hibbins, R. E.; Fossey, S. J.; Moles, J. R.; Sarre, P. J. *Astrophys. J*. **1998**, 495, 941.

[28] Balaban A. T.; Harary, F.; *Tetrahedron*, **1967**, 24, 2505.

[29] Caporossi, G.; Hansen, P. *J. Chem. Inf. Comput. Sci*. **1998**, 38, 610.

[30] Tošić, R.; Masŭlović, D.; Stojmenović, I.; Brunvoll, J.; Cyvin, S. J.; Cyvin, B. J. *J. Chem. Inf. Comput. Sci*. **1995**, 35, 181.

[31] Henson, R. A.; Windlinx, K. J.; Wiswesser, W. J. *Comput. Biomed. Res*. **1995**, 8, 53.

[32] Nikolić, S.; Trinajstić, N.; Knop, J. V.; Müller, W. R.; Szymanski, K. *J. Math. Chem*. **1990**, 4, 357.

[33] Balasubramanian, K.; Kaufman, J. J.; Koski, W. S.; Balaban, A. T. *J. Comput. Chem*. **1980**, 1, 149.

[34] Weisman, J. L.; Timothy, J. L.; Head-Cordon, M. *Spectrochimica Acta A*, **2001**, 57, 931.

[35] Hudgins, D. M.; Bauschlicher, C. W., Jr.; Allamandola, L. J. *Spectrochimica Acta A*, **2001**, 57, 907.

[36] Ohno, K.; Takahashi, R.; Yamada, M.; Isogai, Y. *Int. Elec. J. Mol. Des*. **2002**, 1, 636.

[37] Dewar, M. J. S.; Thiel, W. *J. Amer. Chem. Soc*. **1977**, 99, 4499.

[38] Davis, L. P. *J. Comp. Chem*. **1981**, 2, 433.

[39] Dewar, M. J. S.; Zoebisch, E. G.; Healy, E. F. *J. Amer. Chem. Soc*. **1985**, 107, 3902.

[40] Dewar, M. J. S.; Reynolds, C. H. *J. Comp. Chem*. **1986**, 2, 140.

[41] Ridley, J. E.; Zerner, M. C. *Theo. Chim. Acta*. **1973**, 32, 111.

[42] Zerner, M. C.; Lowe, G. H.; Kirchner, R. F.; Mueller-Westerhoff, U. T. *J. Am. Chem. Soc*. **1980**, 102, 589.

[43] Zerner, M. C. In *Reviews of Computational Chemistry*; Lipkowitz, K. B.; Boyd, D. B., Ed.; VCH Publishing: New York, 1991; Vol. 2, 313.

[44] Anderson, D. Z.; Frisch, J. C.; Masser, C. S. *Appl. Optics* **1984**, 23, 1238.

[45] Okruss, M.; Müller, R.; Hese, A. *J. Molecu. Spec*. **1999**, 193, 293.

[46] Hollas J. M.; Thakur, S. N. *Molecular Physics*, **1971**, 22, 203.

[47] Beck, S. M.; Powers, D. E.; Hopkins, J. B.; Smalley, R. E. *J. Chem. Phys*. **1980**, 73, 2019.

[48] Baskin J. S.; Zewail, A. H. *J. Phys. Chem*. **1989**, 93, 5701.

[49] Warren, J. A.; Hayes, J. M.; Small, G. *J. Chem. Phys*. **1986**, 102, 323.

[50] van Herpen, W. M.; Meerts, W. L.; Dymanus, A. *J. Chem. Phys*. **1987**, 87, 182.

[51] Ohta N.; Baba, H. *Chem. Phys. Lett*. **1987**, 133, 222.

[52] Hoheisel, G.; Heinecke, E.; Hese, A. *Chem. Phys. Lett*. **2003**, 373, 416.

[53] Andrews, P. M.; Pryor, B. A.; Berger, M. B.; Palmer, P. M.; Topp, M. R. *J. Phys. Chem. A*, **1997**, 101, 6222.

[54] Bermudez G.; Chan, I. Y. *J. Phys. Chem*. **1986**, 90, 5029.

[55] Müller, W. R.; Szymanski, K.; Knop, J. V. *J. Comput. Chem.* **11**, 223 (1990).

[56] Zhao, L.; Lian, R.; Shkrob, I. A.; Crowell, R. A.; Pommeret, S.; Chronoster, E. L.; Liu, A. D.; Trifunac, A. D. *J. Phys. Chem. A* **2004**, 108, 25.

[57] Savage, B. D.; Drake, J. F.; Budich, W.; Bohlin, R. C. *Astrophys. J*. **1977**, 216, 291.

[58] Salama, F.; Bakes, E. L. O.; Allamandola, L. J.; Tielens, A. G. G. M. *Astrophys. J*. **1996**, 458, 621.

[59] Gross, E. K. U.; Dobson, J. F.; Petersilka, M. *Top. Curr. Chem*. **1996**, 181, 81.

[60] Bauernschmitt, R.; Ahlrichs, R. *Chem. Phys. Lett*. **1996**, 256, 454.

[61] Becke, A. D. *J. Chem. Phys*. **1993**, 98, 5648.

[62] Lee, C.; Yang W.; Parr, R. G. *Phys. Rev*. **1988**, B37, 785.

[63] Miehlich, B.; Savin, A.; Stoll H.; Preuss, H. *Chem. Phys. Lett*. **1989**, 157, 200.

[64] Parac M.; Grimme, S. *Chem. Phys. Lett*. **2003**, 292, 11.

[65] Sonnentrucker, P.; Foing, B. H.; Ehrenfreund, P. *Adv. Space Res*. **1999**, 24, 519.

[66] Sonnentrucker, P.; Cami, J.; Ehrenfreund, P.; Foing, B. H. *Astron. Astrophys*. **1997**, 327, 1215.

[67] Mulas, G. *Astron. Astrophys*. **1998**, 338 243.

[68] Rouan, D.; Léger, A.; Le Coupanec, P. *Astron. Astrophys*. **1997**, 324, 661.

[69] Head-Gordon, M.; Rico, R. J.; Oumi, M.; Lee, T. J. *Chem. Phys. Lett*. **1994**, 219, 21.

[70] E. A. Mangle and M. R. Topp, J. Phys. Chem. **90**, 802 (1986).

[71] J. Wessel and D. S. McClure, Mol. Cryst. Liquid Cryst. **58**, 121 (1980).

[72] A. R. Lacey, E. F. McCoy and I. G. Ross, Chem. Phys. Lett. **21**, 233 (1973).

[73] G. Fischer, Chem. Phys. **4**, 62 (1974).

[74] R. M. Hochstrasser, Accounts Chem. Res. **1**, 266 (1968).